\documentclass{article}
\usepackage{spconf,amsmath,graphicx,hyperref, amssymb}
\usepackage{multirow}
\usepackage[normalem]{ulem}
\usepackage{makecell}
\usepackage{booktabs}

\title{TIP and Polish: Text-Image-Prototype Guided Multi-Modal Generation via Commonality-Discrepancy Modeling and Refinement}
\twoauthors
{Zhiyong Ma\sthanks{Zhiyong Ma performed the work while chasing its PhD at South China University of Technology}, Jiahao Chen\sthanks{Equal Contribution to Zhiyong Ma}, Qingyuan Chuai\sthanks{Corresponding Author}}
{Cao Tu Li (Guangzhou) Technology Co., Ltd\\
Research and Development Department\\
Guangzhou, Guangdong, China
}
{Zhengping Li}
{Hong Kong Baptist University\\
Faculty of Science\\
Kowloon, Hong Kong}

\begin{document}
%
\maketitle
\begin{abstract}
Multi-modal generation struggles to ensure thematic coherence and style consistency.
Semantically, existing methods suffer from cross-modal mismatch and lack explicit modeling of commonality and discrepancy.
Methods that rely on fine-grained training fail to balance semantic precision with writing style consistency.
These shortcomings lead to suboptimal generation quality.
To tackle these issues, we propose \textbf{\textit{TIPPo}}, a simple yet effective framework with explicit input modeling and comprehensive optimization objectives.
It extracts the input text and images via multi-modal encoder and adapters, then measures the visual prototype.
\textbf{T}extual, \textbf{I}mage, and \textbf{P}rototype signals are then fed to our proposed Dual Alignment Attention and Difference Operator modules before language model decoding.
The proposed \textbf{Po}lishPPO reinforces the style consistency, while the unsupervised contrastive learning during SFT mitigates inter-sample representation collapse.
Experimental results demonstrate the promising performance of \textbf{\textit{TIPPo}} in automatic evaluation and LLM-based criteria for creativity and semantic consistency.
\end{abstract}


\section{Introduction}
\label{sec:intro}
Synthesizing content from multi-modal input signals (e.g., topic and image), Multi-Modal Generation (MMG) enables applications like creative writing, image captioning, and story telling \cite{Nazarieh_Feng_Awais_Wang_Kittler_2024}.
The core pursuit of this field is ensuring \textbf{thematic coherence} and \textbf{stylistic consistency} of the generated results.

For \textbf{thematic coherence} \cite{MMT,SE-TRDF}, current methods face three semantic challenges: (1)~cross-modal mismatch, (2)~neglected measurement of intra-modal discrepancy, and (3)~insufficient mining of both inter- and intra-modal commonality. 
To narrow the cross-modal semantic gaps, most MMG methods adopt pre-trained multi-modal encoder (MME) to extract features from inputs, and equip themselves with pre-trained language model decoders for basic generation ability \cite{mPLUG-Owl,qi2024visionlanguagetasksbenefitlarge}.
Building upon this paradigm, existing methods directly inject textual and visual signals into downstream decoders \cite{StoryLLaVA,mm-cot}.
However, intra-modal semantic variations are ubiquitous and non-negligible, and such unregulated usage of inputs increases the risk of semantic drift from the core theme during generation.
Concurrently, although multi-modal inputs contain rich semantic knowledge (e.g., associations within images, correlations between text and image), the simple integration fails to fully leverage such richness, instead leading to marginal improvements in MMG \cite{HARN,Bie_Yang_Zhou_Ghanem_Zhang_Yao_Wu_Holmes_Golnari_Clifton_et_al}.
Thus, we argue that explicit modeling of semantic discrepancy and commonality is critical to achieving thematic coherence.

For \textbf{stylistic consistency}, the optimization objective should be further associated with the literal and semantic meaning of the method's overall output \cite{FlexMuse, PPO}.
In general, token-level training tasks like the Next Token Prediction (NTP) are widely used in MMG to ensure the semantic precision \cite{HARN}.
However, these fine-grained objectives lack an overview of writing style and struggle to maintain the stylistic unification throughout the entire output \cite{DPO, SC2}.
Therefore, we contend that comprehensive objectives are crucial for polishing the quality of MMG.

To address aforementioned challenges, we propose TIPPo, a Text, Image, and visual Prototype-guided framework that explicitly models the semantic commonality and discrepancy, while reinforcing the generation quality with innovative PolishPPO and unsupervised contrastive learning.
Specifically, with the prototype intuitively carrying intra-modal semantic commonality, two modal-aware adapters are used for downstream task adaptation.
In addition, we design an Dual Alignment Attention module to augment image signals guided by inter- and intra-modal semantic commonality.
Meanwhile, the Difference Operator is designed to model and highlight the intra-modal discrepancy, thus alleviating detail lost.
Finally, the textual signal, enhanced image signals, and differential signal are concatenated and injected into the language model for decoding natural language results. 

To improve the generation quality, we introduce PolishPPO to provide hierarchical supervision (literal-level and article-semantic-level) that explicitly guides style consistency while preserving semantic precision, which is tough for the method supervised by fine-grained optimization objectives.
Besides, the unsupervised contrast is used with supervised fine-tuning (SFT).
It takes the prototype as the positive anchor of the image signal while setting the other samples within the same batch as the negative anchors, to learn and extract more discriminative features under semantic constrain, preventing the representation collapse.

Our key contributions are: (1) Proposing a MMG framework \textbf{\textit{TIPPo}} that explicitly models semantic commonality/discrepancy to ensure thematic coherence, and integrates comprehensive optimization to enhance stylistic consistency; (2) Introducing PolishPPO and prototype-driven contrastive learning strategies to balance semantic precision with stylistic consistency and mitigate representation collapse.




\section{Proposed Method}
\label{sec:method}
\begin{figure}[ht]
\centering
    \includegraphics[width=0.44\textwidth]{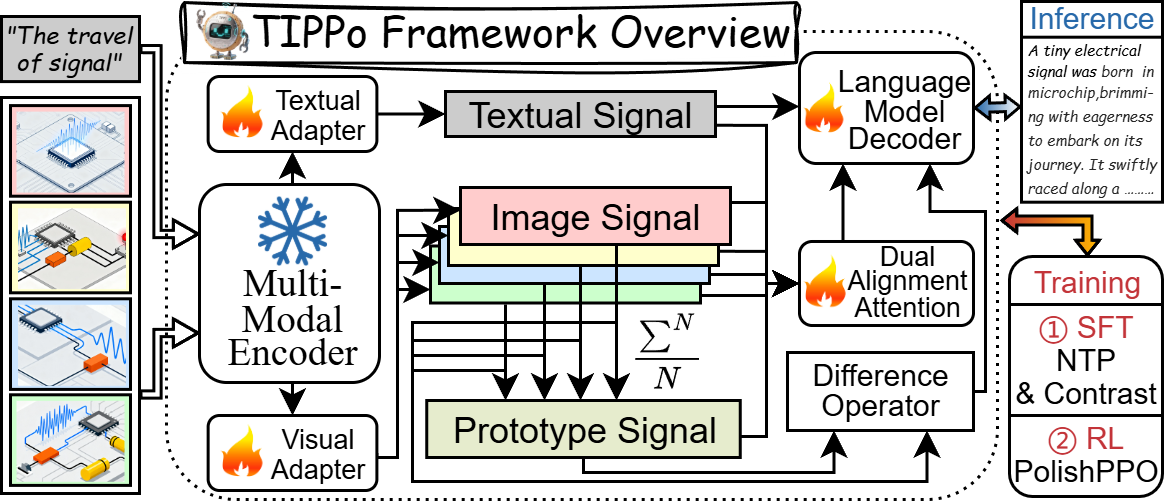}
    \caption{The overview of TIPPo framework}
    \label{fig:TIPPo}
\end{figure}

We propose TIPPo, a multimodal generation framework that captures inter-modal commonalities and intra-modal differences, with various training objectives, as shown in Fig.\ref{fig:TIPPo}. 
Firstly, the given text $\mathcal{T}$ and $N$ associated images $\mathcal{I} = \{\mathcal{I}_1, \dots, \mathcal{I}_N\}$, TIPPo extracts the corresponding signals by a frozen MME $\phi$ and trainable two adapters, $\pi_T$ and $\pi_V$, where the textual signal $S^T = \pi_T(\phi(\mathcal{T}))$, the image signals $S^I = \{S_i^I=\pi_T^I(\phi(\mathcal{I}_i))~|~i=1,\dots,N\}$, and the prototype signal $S^P = \frac{1}{N} \sum_{i}^{N} {S_i^I}$.
Leveraging $S^T$, $S^I$ and $S^P$, the Dual Alignment Attention module augments each image signal, and the corresponding result $S_i^A$ and $S_A$ are as Eq.\ref{eq: S_A}-\ref{eq: S_i_A}.
Then $S^I$ and $S^P$ are used to get the differential signal of $i$-th image $S^D_i$ and the corresponding set $S^D$ by the Difference Operator as Eq. \ref{eq: S_D}-\ref{eq: S_i_D}.
Finally, textual signal $S^T$, augmented signals $S^A$ and differential signal $S^D$ are concatenated to get the fusion signal $S^{F}=\left\{S_{i}^{F}=\left[S^{T};S_{i}^{A};S_{i}^{D}\right]~|~i=1,2,\cdots,N\right\}$.
The language model decoder $LM(\cdot)$ generate the result $y = LM(S^F)$.
During training, we first conduct SFT with NTP and contrastive constrain (see Eq. \ref{eq: contrastive}) as: $\mathcal{L}_{SFT} =  \mathcal{L}_{NTP} + \mathcal{L}_{C}$, then use PolishPPO to further optimize.
Details are as follow.

\subsection{Dual Alignment and Differential Signal}
\label{subsec:DADS}
Existing methods demonstrated the effectiveness about the alignment process in capturing the semantic commonality \cite{Nazarieh_Feng_Awais_Wang_Kittler_2024,StoryLLaVA}.
To leverage the inter- and intra-modal semantic commonality to enhance image signals, the Dual Alignment Attention (DAA) module is proposed.
It builds the text-image alignment attention map and the prototype-image alignment attention map, where the first map tries to mine the inter-modal semantic commonality, while the second map focuses on measuring intra-modal semantic commonality.
Two dual maps jointly enhance image signals for better MMG.

Specifically, it sets the textual query $Q^T = W^{TQ} \odot S^T$, prototype query $Q^P = W^{PQ} \odot S^P$, text-image key $K^T = \{K_i^T = W^{TIK} \odot S_i^I~|~i=1,\dots,N\}$, prototype-image key $K^P = \{K_i^P = W^{PIK} \odot S_i^I~|~i=1,\dots,N\}$ and visual values $V^I = \{V_i^I = W^V \odot S_i^I~|~i=1,\dots,N\}$, where all the $W^*$ are trainable parameters.
Denoting $\sigma(\cdot)$ as the softmax operator, the augmented image signal $S^A$ is constructed by:
\begin{equation}
\label{eq: S_A}
S^A = \{S_i^A~|~i=1,\dots,N\}
\end{equation}
\begin{equation}
\label{eq: S_i_A}
S_{i}^{A} = \bigg[\lambda_{1} \cdot \sigma\left(\frac{Q^{P}K^{P}}{\sqrt{d_{k}}}\right) +\lambda_{2} \cdot\sigma\left(\frac{Q^{T}K^{T}}{\sqrt{d_{k}}}\right) \bigg]V_{i}^{I}.
\end{equation}
Notably, $\lambda_1$ and $\lambda_2$ in Eq. $\ref{eq: S_i_A}$ are tied to the training stage, with $\lambda_1=m/M$ and $\lambda_2=1-\lambda_1$ ($m$: current step, $M$: total steps).
This dual-weight scheduling balances learning generalizable prototype-guided semantics in the early phase and refining personalized inter-modal details as training advances, tackling the generality-specificity trade-off in MMG.
As $\lambda_1$ rises, it gradually emphasizes prototype-grounded semantic detail augmentation for images, while $\lambda_2$ sustains explicit inter-modal commonality modeling and attention to personalized fine-grained details.

In addition, to fully exploit the unique visual detail, we propose the Difference Operator (DO).
Existing methods predominantly focus on capturing commonalities (e.g., shared semantic concepts) \cite{Bie_Yang_Zhou_Ghanem_Zhang_Yao_Wu_Holmes_Golnari_Clifton_et_al,mm-cot}. 
However, intra-modal discrepancies are also critical for generating fine-grained content. 
For instance, in scenes with multiple similar objects, intra-modal discrepancies in image features like subtle shape or texture variations are intuitively crucial for distinguishing each object precisely.
Thus, the Difference Operator targets this unexploited dimension by measuring and refining intra-modal differential information to obtain the differential signal $S^D$, where $\pi_D$ represents a trainable linear layer:
\begin{equation}
\label{eq: S_D}
S^D = \{S_i^D~|~i=1,\dots,N\}
\end{equation}
\begin{equation}
\label{eq: S_i_D}
S_{i}^{D}=(S_{i}^{I}-S^{P})+\pi_D(S_{i}^{I}-S^{P}).
\end{equation}
Notably, the residual design in Eq. \ref{eq: S_i_D} enables flexible control over the strength of $S^D$. 
When the differential information is not beneficial or even detrimental to the downstream task (e.g., noisy details), the $\pi_D$ can adaptively reduce the influence. 
This mechanism allows our framework to neglect or even degrade inappropriate differential information, ensuring that only useful differential information contributes to the final generation, enhancing the robustness and adaptability.

\subsection{Contrastive Constraint and PolishPPO}
To mitigate representation collapse we design an unsupervised contrastive constraint in SFT that brings the representation of each image closer to its corresponding prototype while distinguishing from others. 
In other words, we take the prototype signal as the positive anchor of each image while taking the other images as the negative set in InfoNCE \cite{SimCLR}:  
\begin{equation}
\label{eq: contrastive}
\mathcal{L}_{C}=-\frac{1}{N}\sum_{i=1}^{N}\log\left(\frac{\exp(cos(S_i^I,S_i^P))}{\sum_{j}^{N}\exp(cos(S_i^I,S_j^I))}\right).
\end{equation}
In cases where only one corresponding image exists (like one image pair for one or no textual instruction in image captioning), the prototype can be measured as the average of the image signals from the same batch, and negative samples are the other images within the same batch.    

In addition, the fine-grained optimization objective is hard to explicitly guide style consistency
while preserving semantic precision.
PolishPPO is proposed to address this dilemma by providing literal-level and article-semantic-level supervision.
It adjusts the \textbf{cumulative discounted return $G_m$}, and the \textbf{normalized advantage estimation $\hat{A}$} in PPO \cite{PPO}:
\begin{equation}
\label{eq: advantage}
\hat{A}:=\frac{G_m-\mu_m}{\delta_m+\epsilon}, ~~\epsilon:=1e-8,
\end{equation}
where $\mu_m$ and $\delta_m$ are the expectation and variance over the last $m$ iterations.
$G_m=\gamma^{m}R_{literal}+\gamma^{M-m}R_{semantic}$, where $\gamma$ is a scalar hyperparameter set as 0.95, with its exponents determined by current and total training steps ($m$ and $M$).
Denote the ground-truth of the generation result $y$ as $Y$.
We obtain their embedding $emd_{Y}$ and $emd_{y}$ via GLM \cite{ChatGLM}.
Denotes the $LD(\cdot,\cdot)$ as the Levenshtein distance operator, and $cos$ as the cosine operator same in Eq. \ref{eq: contrastive}
Then literal-level reward $R_{literal}$ and the semantic-level rewar $R_{semantic}$ used in $G_m$ are respectively obtained as:
\begin{equation}
\label{eq: literal_reward}
R_{literal}=1-LD(Y,y),
\end{equation}
\begin{equation}
\label{eq: semantic_reward}
R_{semantic}=cos(emd_{Y},emd_{y}).
\end{equation}

\section{Experiments}
\label{sec:experiments}
\subsection{Experimental settings}
TIPPo is evaluated on two MMG datasets. 
The first is \textit{ArtMuse} \cite{FlexMuse}, a creative writing dataset with 100 articles, each contains a textual title, 3–5 associated images, and a reference result.
The second is \textit{COCO-CN} \cite{COCO-CN}, a mage captioning dataset in Chinese (built on \textit{MS-COCO}) with 22k manual sentences and 5k translated sentences.
We adopt two test beds upon automatic evaluation and LLM-based criteria measure the generation quality.
For automatic evaluation, we employ metrics including Rouge-L, METEOR, CIDEr, and BertScore following \cite{li2023associative,su2024glocal}. 
Considering the limitations of these metrics in holistically, we further use LLM-based criteria for creativity and semantic consistency following \cite{FlexMuse}, which assesses results from 10 perspectives.
We initialize the MME and the Language Model Decoder from \textit{OFA-Sys/chinese-clip-vit-base-patch16} and \textit{langboat/mengzi-t5-base}.
During SFT, TIPPo trains 50K iterations with learning rate $\eta_{SFT}=5e-5$ verified from \{3e-5, 4e-5, 5e-5, 6e-5, 7e-5\}. 
The PolishPPO executes 20K iterations with $\eta_{PPO}=1e-5$ verified from \{6e-6, 9e-6, 1e-5, 3e-5, 5e-5\}.

\subsection{Experimental Results}
Overall, TIPPo achieves promising performance in different aspects.
Specifically, as shown in Table \ref{tab:ArtMuse}, TIPPo shows decent performance within 3 metrics on \textit{ArtMUSE}, where METEOR, ROUGE-L and BertScore respectively increase 4.56, 8.1 and 0.06 relative to the second best method.
\begin{table}[htbp]
  \centering
  \caption{Automatic evaluation on \textit{ArtMUSE}. 
  The best and the second results are in \textbf{bold} and \uline{underline} respectively.}
  \renewcommand{\arraystretch}{1} 
  \begin{tabular}{lccc}
    \toprule
    methods & METEOR & ROUGE-L & BertScore \\
    \midrule
    mm-cot \cite{mm-cot}       & 1.64  & 12.02  & 0.43 \\
    mPLUG-Owl \cite{mPLUG-Owl}    & 4.48  & 14.21  & 0.63 \\
    LaDic \cite{LaDiC}        & 7.31  & 12.13  & 0.55 \\
    DOC \cite{DOC}          & 1.99  & 8.02   & 0.65 \\
    GLM-4V \cite{ChatGLM}       & 1.41  & 9.34   & 0.66 \\
    Qwen2.5-v1 \cite{Qwen2.5-VL}   & 1.28  & 9.37   & 0.65 \\
    GPT-4o \cite{GPT-4o}       & 3.53  & 9.41   & 0.66 \\
    FlexMUSE \cite{FlexMuse}     & \uline{23.32} & \uline{30.18}  & \uline{0.71} \\
    \midrule
    \textbf{TIPPo}        & \textbf{27.88} & \textbf{38.28}  & \textbf{0.77} \\
    \bottomrule
  \end{tabular}
  \label{tab:ArtMuse}
\end{table}
\begin{figure*}[htbp]
   \centering
    \includegraphics[width=0.9\textwidth, height=2.7cm]{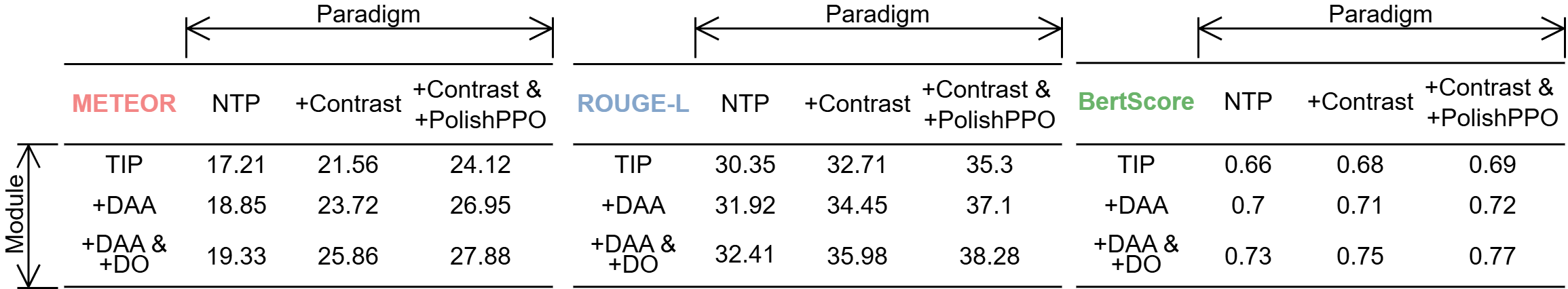}
    \caption{Effectiveness of individual modules or methods in TIPPo evaluated by automatic metric on ArtMUSE.}
    \label{fig:Ablation}
\end{figure*}

Besides, TIPPo achieves the highest performance on \textit{COCO-CN} in METEOR and ROUGE-L, exceeding HARN \cite{HARN} by 21.22 and 3.96, respectively, as shown in Table \ref{tab:COCO-CN}. 
However, its CIDEr remains lower than HARN. 
This performance gap might be caused by CIDEr prioritizes the coverage of high-frequency key words in reference texts, while TIPPo focuses more on the overall semantic consistency and logical coherence of generated descriptions rather than deliberately matching such high-frequency local details.

\begin{figure}[htbp]
    \centering
    \includegraphics[width=0.8\linewidth, height=6cm]{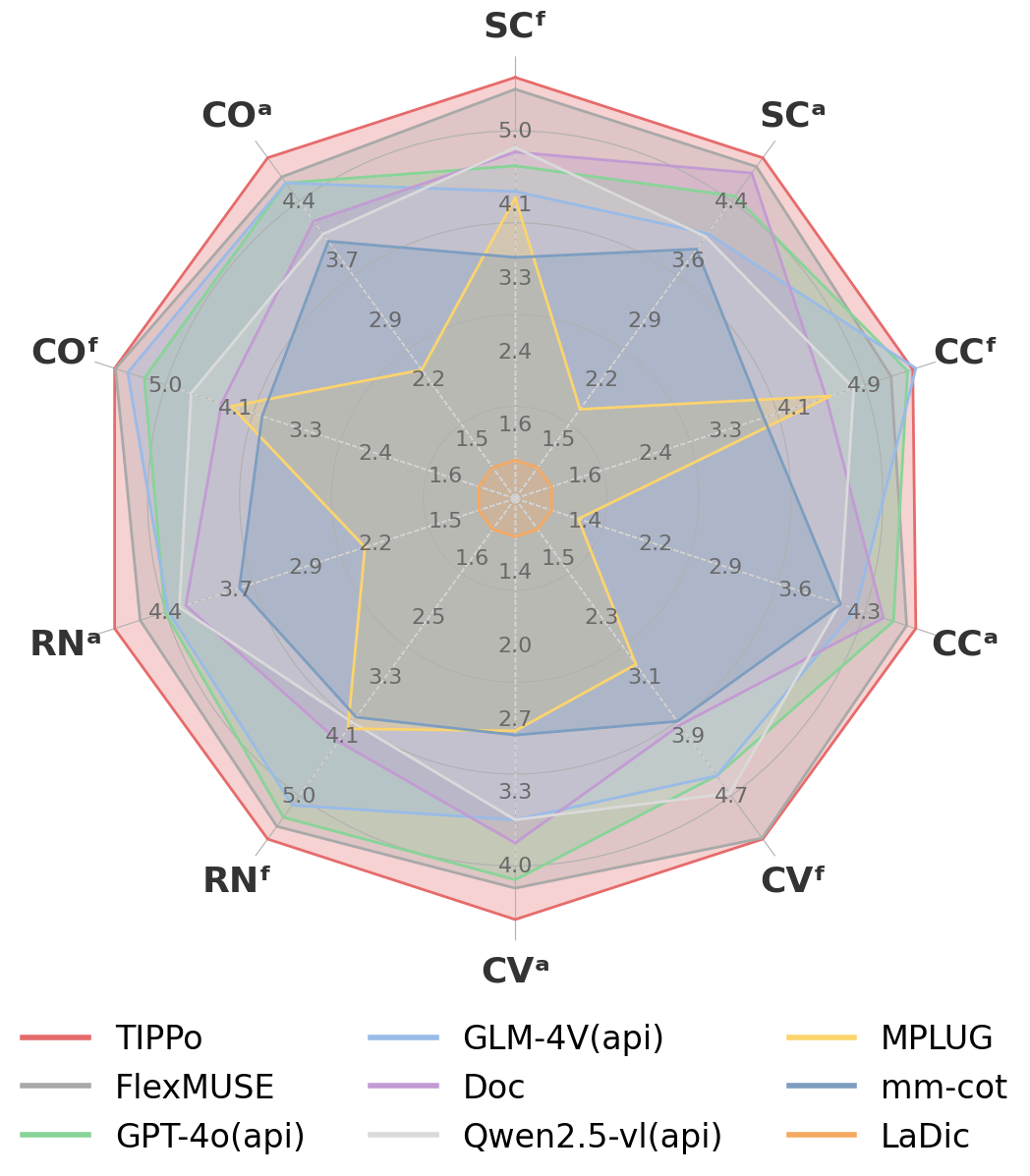}
    \caption{
    LLM evaluation of different methods on ArtMUSE.
    }
    \label{fig:LLMscore}
\end{figure}

As shown in Fig. \ref{fig:LLMscore}, TIPPo demonstrates promising results in LLM evaluation.
Specifically, TIPPo outperforms all comparison methods in $CC^f$, $CC^a$, $CO^f$, and $CO^a$ (defined in \cite{FlexMuse}), reflecting its superiority in addressing thematic coherence issues. 
The performance about $SC^f$ and $SC^a$ (also defined in \cite{FlexMuse}) highlight the effectiveness in solving the stylistic consistency challenge.

\subsection{Ablation study}

We record the ablation results in Fig. \ref{fig:Ablation}, evaluating the contributions of DAA, DO and training strategies. 
We denotes TIP as a baseline method which ablate the DAA and DO from TIPPo.
Module-wise (vertical) comparisons reveal consistent performance enhancements with the incremental integration of components across all optimization paradigms.
Paradigm-wise (horizontal) comparisons demonstrate progressive enhancements from advanced optimization objectives. 
These results validate the contribution of each design in TIPPo.

\begin{table}[htbp]
  \centering
  \caption{Automatic evaluation on \textit{COCO-CN}. 
  The best and the second results are in \textbf{bold} and \uline{underline} respectively.}
  \renewcommand{\arraystretch}{1} 
  \begin{tabular}{lccc}
    \toprule
    methods & METEOR & ROUGE-L & CIDEr \\
    \midrule
    Meng et al. \cite{meng2022objectcentricunsupervisedimagecaptioning}  & 12.1   & 22.8    & 34.9  \\
    Ukyo et al. \cite{honda-etal-2021-removing}  & 12.8   & 23.4    & 19.7  \\
    HARN \cite{HARN}        & \uline{29}     & \uline{50.3}    & \textbf{89.8}  \\
    \midrule
   \textbf{TIPPo}       & \textbf{50.22}  & \textbf{54.26}   & \uline{59.35} \\
    \bottomrule
  \end{tabular}
  \label{tab:COCO-CN}
\end{table}
\begin{table}[htbp]
    \centering
    \caption{Comparison about the of optimization objectives.} 
    \begin{tabular}{l@{\hspace{2pt}}c@{\hspace{2pt}}c@{\hspace{2pt}}c@{\hspace{2pt}}c}
        \toprule
        ArtMuse   &Iterations   & METEOR  & ROUGE-L & BertScore \\
        \midrule
        PolishPPO Only & 20k& 3.26  & 4.57  &  0.57 \\
        & 50k& 5.64  & 8.97  &  0.58 \\
        & 70k& 6.21  & 11.24  &  0.60 \\
        & 80k& 6.06  & 11.14  &  0.60 \\
        & 90k& 6.11  & 11.17  &  0.59 \\
        \midrule
        NTP Only & 30k & 8.34   & 15.39   & 0.62      \\
        & 40k& 8.77    & 16.98    & 0.62     \\
        & 50k& 10.82   & 18.43   & 0.62      \\
        & 60k& 10.23  & 18.14  &  0.62 \\
        & 70k& 9.98  & 16.24  &  0.61 \\
        \midrule
        Both & 50+20k & 17.53   & 27.89   & 0.64      \\
        \bottomrule
    \end{tabular}
    \label{tab:PolishPPO} 
\end{table}

To further validate the effectiveness of PolishPPO, we conducted additional verifications with different optimization objectives using the TIP without prototype input, as shown in Table \ref{tab:PolishPPO}. 
When only NTP is employed, the performance in three criterias is almost at the peak once the iteration reaches 50k.
In comparison, continuing optimization with PolishPPO following 50k steps of NTP (last line records) show effective improvement in model performance.
Moreover, we also set up several test beds that training the baseline only with PolishPPO.
As expected, the performance of the baseline gradually improved with iteration, and approached the optimal result with 70k iterations.
The above validations indicate that PolishPPO, as a complementary refinement step to SFT (NTP), can effectively improve the overall quality of MMG.
\section{Conclusion}
\label{sec:conclusion}

In this work, we propose a text-image-prototype guided multi-modal generation framework, namely TIPPo.
To ensure thematic coherence, TIPPo integrates the Dual Alignment Attention to augment image signals guided by inter- and intra-modal semantic commonality.
Additionally, a Difference Operator is integrated to model and highlight the intra-modal discrepancy, thus alleviating detail lost. 
Furthermore, to improve the generation quality, we introduce PolishPPO to provide hierarchical s upervision that  explicitly guides style consistency.
Enabled by its well-designed modules and training strategies, TIPPo achieves promising performance on ArtMuse and COCO-CN datasets. 
Future work will explore extending TIPPo to more complex multi-modal scenarios (e.g., video-text generation) and optimizing its efficiency for real-time applications.

\vfill\pagebreak


\bibliographystyle{IEEEbib}
\bibliography{refs}

\end{document}